\documentclass{ws-procs9x6}

\begin{document}

\title{Indirect Searches for Kaluza-Klein Dark Matter\footnote{\uppercase{B}ased on work with \uppercase{G}raham \uppercase{K}ribs and \uppercase{J}oseph \uppercase{S}ilk.}}

\author{Dan Hooper\footnote{\uppercase{S}upported by the \uppercase{L}everhulme Trust}}

\address{University of Oxford, \\
Denys Wilkinson Building, Keble Road, \\ 
Oxford, OX1-3RH, UK\\ 
E-mail: hooper@astro.ox.ac.uk}

\maketitle

\abstracts{
In this talk, we discuss the potential for the indirect detection of Kaluza-Klein dark matter using neutrino telescopes and cosmic positron experiments. We find that future kilometer-scale neutrino telescopes, such as IceCube, as well as future experiments capable of measuring the cosmic positron spectrum, such as PAMELA and AMS-02, will be quite sensitive to this scenario. Current data from the HEAT experiment can also be explained by the presence of Kaluza-Klein dark matter in the Galactic halo.}

\section{Introduction}

Although there exists an enormous body of evidence for the existence of dark matter, its identity remains unknown\cite{review}. Weakly Interacting Massive Particles (WIMPs) are, perhaps, the most well motivated class of candidates for dark matter. Among these, the lightest neutralino in models of supersymmetry is the most popular. 

Models with extra spatial dimensions can provide an alternative candidate for dark matter, however. In particular, in models in which all of the Standard Models fields are free to propagate in the bulk, called universal extra dimensions, the Lightest Kalzua-Klein Particle (LKP) may be stable and a potentially viable dark matter candidate\cite{taitservant,fengmatchevcheng}. 

The most natural LKP is the first Kaluza-Klein excitation of the hypercharge gauge boson, $B^{(1)}$. We simply refer to this state as Kaluza-Klein Dark Matter (or KKDM) throughout this talk. Previous studies of KKDM have found that the relic density predicted for such a state would naturally coincide with the measurements of WMAP for masses near about $m_{\rm{LKP}}\simeq 800$ GeV if no other Kaluza-Klein states participate in the freeze-out process. If other states are light enough to significantly effect this process, however, the LKP can be substantially lighter\cite{taitservant}.

For the purposes of indirect detection, KKDM has several interesting phenomenological features. First, approximately $60\%$ of their annihilations are to charged lepton pairs ($20\%$ to each generation). $33\%$ of annihilations produce pairs of up-type quarks and $3.6\%$ produce neutrino pairs. The remaining fraction generate down type quarks and Higgs bosons. This is in stark contrast to neutralinos which do not annihilate efficiently to neutrinos, positrons, muons or other light fermions. Second, the total annihilation cross section for KKDM is given by
\begin{equation}
<\sigma v> = \frac{95 g_1^4}{324 \pi m^2_{LKP}} \simeq \frac{1.7 \times 10^{-26} \, \rm{cm^3}/\rm{s}}{m^2_{LKP}(\rm{TeV})}. 
\end{equation}
Notice that this consists entirely of an $a$-term in the expansion, $<\sigma v> =a + bv^2 + \mathcal{O}$$(v^4)$, thus the low velocity cross section is naturally the maximum possible for a thermal relic.

\section{Indirect Detection with Neutrino Telescopes}

Dark matter particles travelling through the Galactic halo can occasionally scatter and become trapped in deep gravitational wells, such as the Sun or Earth. Within these bodies, they accumulate and their annihilation rate is enhanced, potentially providing an observable flux of high-energy neutrinos\cite{indirectneutrino}. 

The capture rate of KKDM particles in the Sun is given by\cite{kkneu}
\begin{equation}
C^{\odot} \simeq 3.35 \times 10^{18} \rm{s}^{-1} \bigg(\frac{\sigma_{\rm{H, SD}}}{10^{-6}\, \rm{pb}}\bigg) \bigg(\frac{1000 \, \rm{GeV}}{m_{\rm{LKP}}}\bigg)^2,
\end{equation}
where $\sigma_{\rm{H, SD}}$ is the spin-dependent, elastic scattering cross section of KKDM off of hydrogen. This expression assumes a local dark matter density of 0.3 GeV/cm$^3$ and a RMS velocity of 270 km/s. The elastic scattering cross section is given by\cite{taitservantdirect}
\begin{equation}
\sigma_{\mathrm{H,SD}} = \frac{g'^4 m_p^2}{648 \pi m_{\rm{LKP}}^4 r_{q^{(1)}}^2} 
\left( 4 \Delta_u^p + \Delta_d^p + \Delta_s^p \right)^2,
\end{equation}
where $r_{q^{(1)}}=(m_{q^{(1)}}-m_{\rm{LKP}})/m_{\rm{LKP}}$ is the fractional shift of the Kaluza-Klein quark masses over the LKP mass and the $\Delta^p_q$'s parameterize the fraction of spin carried by 
a constituent quark $q$. Inserting numerical values for the $\Delta^p_q$'s, we get 
\begin{equation}
\sigma_{\rm{H, SD}} \simeq 0.9 \times 10^{-6} \, \mathrm{pb} 
\left( \frac{1000 \, \mathrm{GeV}}{m_{\rm{LKP}}} \right)^4 
\left( \frac{0.14}{r_{q^{(1)}}} \right)^2 \; .
\end{equation}
For the annihilation and elastic scattering cross sections of KKDM, the annihilation rate in the Sun should reach (or nearly reach) equilibrium with the capture rate. These annihilations can produce neutrinos directly or in the decays of tau leptons or quarks\cite{kkneu}.

\begin{figure}[ht]
\centerline{\epsfxsize=3.0in\epsfbox{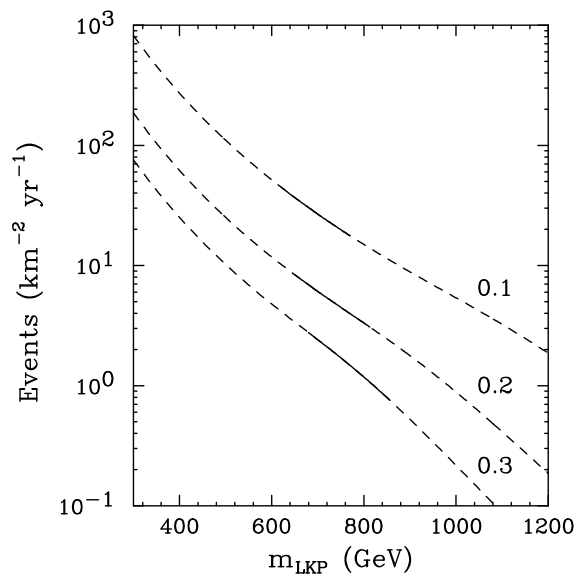}}   
\caption{The rate of muon-induced neutrinos above 50 GeV predicted in a kilometer-scale neutrino telescope, such as IceCube. Curves are shown for Kaluza-Klein quarks 10$\%$, 20$\%$ and 30$\%$ heavier than the LKP \label{neutrinorate}.}
\end{figure}

Muon neutrinos which reach the Earth from the Sun can scatter in charged current interactions with nucleons to produce high-energy muons. These muons produce observable ``tracks'' as they propagate through the medium of a neutrino telescope, such as the Antarctic Ice of the IceCube experiment. The rate of muon tracks generated in a kilometer-scale neutrino telescope from KKDM annihilations in the Sun in shown in figure~\ref{neutrinorate}. Notice that these results depend strongly on the LKP's mass and the mass of the Kaluza-Klein quarks. Calculations of the radiative corrections to the Kaluza-Klein spectrum estimate values of $r_{q^{(1)}}$ roughly in the range of 0.1 to 0.2\cite{radiative}. For a 800 GeV LKP, about 5 to 50 events per year are predicted in IceCube over this range. For a lighter LKP of 500-600 GeV, up to 100 events could be observed.

\section{Indirect Detection with Positron Experiments}

Dark matter annihilating in the Galactic halo can produce several potentially observable species of cosmic rays, including gamma-rays, anti-protons, anti-deuterons and positrons. Kaluza-Klein Dark Matter has characteristics which are favorable for detection with cosmic positrons: a large low-velocity annihilation cross section and a large fraction of annihilations which produce energetic positrons (such as the modes $e^+ e^-$, $\mu^+ \mu^-$ and $\tau^+ \tau^-$). 

To calculate the observed positron spectrum, the positrons injected in dark matter annihilations must be propagated through the Galactic magnetic fields, including scattering with starlight and the CMB. These effects are taken into account by solving the diffusion-loss equation:
\begin{equation}
\frac{\partial}{\partial t}\frac{dn_{e^{+}}}{dE_{e^{+}}} = \vec{\bigtriangledown} \cdot \bigg[K(E_{e^{+}},\vec{x})  \vec{\bigtriangledown} \frac{dn_{e^{+}}}{dE_{e^{+}}} \bigg] \frac{\partial}{\partial E_{e^{+}}} \bigg[b(E_{e^{+}},\vec{x})\frac{dn_{e^{+}}}{dE_{e^{+}}}  \bigg] + Q(E_{e^{+}},\vec{x}),
\end{equation}
where $K(E_{e^{+}},\vec{x})$ is the diffusion constant, $b(E_{e^{+}},\vec{x})$ is the energy loss rate and $Q(E_{e^{+}},\vec{x})$ is the source term.

\begin{figure}[ht]
\centerline{\epsfxsize=3.0in\epsfbox{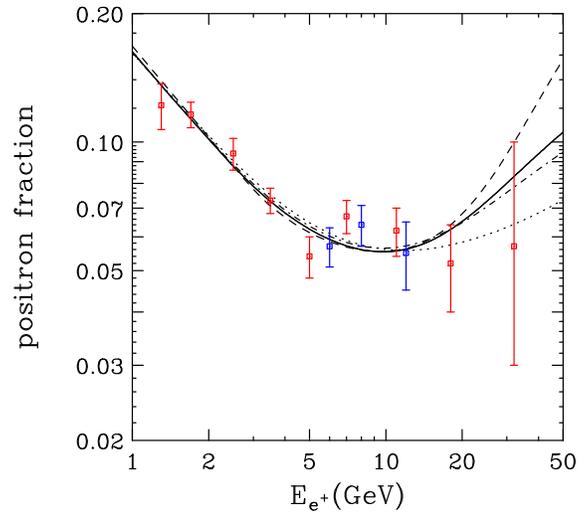}}   
\caption{The ratio of positrons to positrons plus electrons with a contribution from annihilating Kaluza-Klein Dark Matter compared to the HEAT data. Results for several choices of diffusion parameters are shown. The annihilation rate was normalized to the data \label{heat}.}
\end{figure}

The HEAT experiment, during balloon flights in 1994-95 and 2000, observed an excess in the cosmic positron flux when compared to the electron flux. This excess peaks near 8-10 GeV and extends to above 30 GeV where the detector's sensitivity falls off\cite{heat}. It has been suggested that this excess could be the product of dark matter annihilations\cite{positrons}. Neutralino dark matter, which does not annihilate to light fermions, produces positrons inefficiently, however, and therefore requires a very high annihilation rate ({\it i.e.} a clumpy distribution) to account for the excess observed by HEAT\cite{clumps}. KKDM, on the other hand, can produce such positron fluxes more naturally\cite{kkpos}. Rather light LKPs (300-400 GeV) are required to generate the HEAT excess, however, which can only provide the observed relic density if other Kaluza-Klein modes play a very significant role in the freeze-out process\cite{kkpos}.

\begin{figure}[ht]
\centerline{\epsfxsize=3.0in\epsfbox{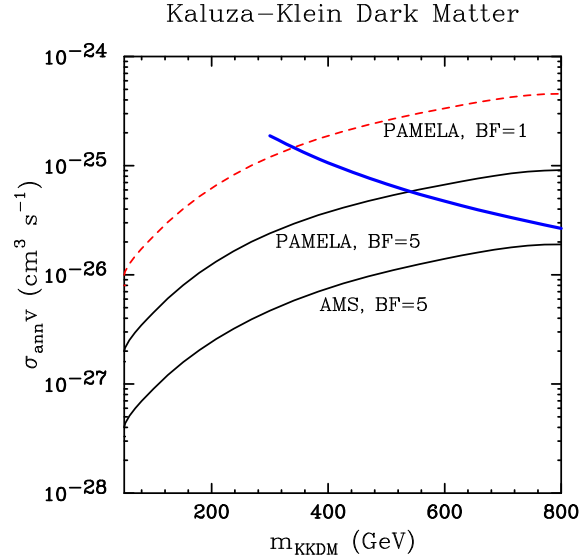}}   
\caption{The sensitivity of PAMELA and AMS-02 to cosmic positrons generated in Kaluza-Klein Dark Matter (KKDM) annihilations. Contours are shown for a completely homogeneous distribution (BF=1) and a mildly clumped distribution (BF=5). The downward sloping curve represents the annihilation cross section predicted for KKDM.    \label{future}}
\end{figure}

Future experiment, such as PAMELA and AMS-02, will be capable of measuring the cosmic positron spectrum with much greater precision and to much higher energies than HEAT\cite{pos2,futurepositron}. These experiments should be very sensitive to the presence of KKDM in our Galaxy\cite{pos2}. In figure~\ref{future}, we show the sensitivity of these experiments to KKDM. 

\section{Summary}

The prospects for the indirect detection of Kaluza-Klein Dark Matter (KKDM) are very promising for both future neutrino telescopes and cosmic positron experiments. Experiments such as IceCube, PAMELA and AMS-02 will provide powerful probes of KKDM in the coming years.


\end{document}